# Einstein's steady-state theory: an abandoned model of the cosmos

Cormac O'Raifeartaigh,[a] Brendan McCann,[a] Werner Nahm[b] and Simon Mitton[c]

[a]*Department of Computing, Maths and Physics, Waterford Institute of Technology, Cork Road, Waterford, Ireland*

[b]*School of Theoretical Physics, Dublin Institute for Advanced Studies, Burlington Road, Dublin 2, Ireland*

[c]*Department of the History and Philosophy of Science, University of Cambridge, Cambridge, United Kingdom*

Author for correspondence: coraifeartaigh@wit.ie

## Abstract

We present a translation and analysis of an unpublished manuscript by Albert Einstein in which he attempted to construct a 'steady-state' model of the universe. The manuscript, which appears to have been written in early 1931, demonstrates that Einstein once explored a cosmic model in which the mean density of matter in an expanding universe is maintained constant by the continuous formation of matter from empty space. This model is very different to previously known Einsteinian models of the cosmos (both static and dynamic) but anticipates the later steady-state cosmology of Hoyle, Bondi and Gold in some ways. We find that Einstein's steady-state model contains a fundamental flaw and suggest that it was abandoned for this reason. We also suggest that he declined to explore a more sophisticated version because he found such theories rather contrived. The manuscript is of historical interest because it reveals that Einstein debated between steady-state and evolving models of the cosmos decades before a similar debate took place in the cosmological community.

## 1. Introduction

In the course of our research into Einstein's cosmology in the early 1930s,[1] we recently came upon an attempted model of the cosmos that he did not publish. This model is set out in a signed, four-page handwritten manuscript entitled *"Zum kosmologischen Problem"* in the Albert Einstein Archives of the Hebrew University of Jerusalem (Einstein 1931a). Until now,[2] the manuscript was understood to be a draft of the paper *"Zum kosmologischen Problem der allgemeinen Relativitätstheorie"*, published by Einstein in the journal *Sitzungsberichte der Königlich Preussischen Akademie der Wissenschaften* (Einstein 1931b). However, on examining the manuscript we found that it features a model of the cosmos strikingly different to the *Sitzungsberichte* paper, despite some similarities in title and opening paragraphs (figure 1).

The manuscript indicates that Einstein once explored a model of an expanding universe in which the density of matter remains constant due to a continuous formation of matter from empty space. This model is very different to other Einsteinian models of the cosmos, such as his early model of a static universe (Einstein 1917) or his later models of a dynamic, evolving universe (Einstein 1931b, Einstein and de Sitter 1932). However, the model anticipates the well-known steady-state theories of Fred Hoyle, Hermann Bondi and Tommy Gold (Hoyle 1948; Bondi and Gold 1948) in some ways.

We present a guided tour of Einstein's manuscript in section 2 of our paper and discuss a likely date for the work in section 3. In section 4, we consider possible reasons for Einstein's abandonment of his steady-state theory, not least a fundamental flaw in the derivation. The model is considered in the context of other steady-state theories in section 5, and in the context of Einstein's philosophy of cosmology in section 6. We provide a translation of the original manuscript in an Appendix by kind permission of the Albert Einstein Archive of the Hebrew University of Jerusalem.

---

[1] See (O'Raifeartaigh and McCann 2014).
[2] This point has been confirmed by John Stachel, the founding editor of the Einstein Papers Project (Stachel 2013).

## 2. A guided tour of the manuscript[3]

The manuscript begins with Einstein recalling the well-known problem of gravitational collapse in a Newtonian universe. This starting point is similar to Einstein's seminal cosmological paper of 1917 (Einstein 1917), although he now includes a reference to the work of Hugo Seeliger:

> *"It is well known that the most important fundamental difficulty that emerges when one asks how the stellar matter fills up space in very large dimensions is that the laws of gravity are not in general consistent with the hypothesis of a finite mean density of matter. Thus, at a time when Newton's theory of gravity was still generally accepted, Seeliger had already modified the Newtonian law by the introduction of a distance function that, for large distances r, diminishes considerably faster than $1/r^2$."*

Einstein points out that a similar problem arises in relativistic models of the cosmos, and recalls his introduction of the cosmological constant to the field equations of relativity to render them consistent with a static universe of constant radius and matter density:

> *"This difficulty also arises in the general theory of relativity. However, I have shown in the past that this can be overcome by the introduction of the so-called "λ–term" to the field equations. The field equations can then be written in the form*
>
> $$\left(R_{ik} - \frac{1}{2}g_{ik}R\right) - \lambda g_{ik} = \kappa T_{ik} \qquad \ldots (1)$$
>
> $$R_{ik} = \Gamma^{\sigma}_{ik,\sigma} - \Gamma^{\sigma}_{i\sigma,k} - \Gamma^{\sigma}_{i\tau}\Gamma^{\tau}_{k\sigma} + \Gamma^{\sigma}_{ik}\Gamma^{\tau}_{\sigma\tau} \qquad \ldots (1a)$$
>
> *At that time, I showed that these equations can be satisfied by a spherical space of constant radius over time, in which matter has a density ρ that is constant over space and time."*

In the next part of the manuscript, Einstein suggests that this static model now seems unlikely:

> *"It has since transpired that this solution is almost certainly ruled out for the theoretical comprehension of space as it really is"*

We note that Einstein dismisses his static model for two separate reasons. First, he comments on the existence of dynamic solutions and notes that his static solution was found to be unstable:

> *"On the one hand, it follows from investigations based on the same equations by [ ] and by Tolman that there also exist spherical solutions*

---
[3] We suggest this section be read in conjunction with our translation of Einstein's manuscript in the Appendix.

> *with a world radius P that is variable over time, and that my solution is not stable with respect to variations of P over time."*

The blank space in the sentence above representing theoreticians other than Tolman who suggested dynamic solutions is puzzling as Einstein was unquestionably aware of the cosmological models of both Friedman and Lemaître (Nussbaumer 2014; Nussbaumer and Bieri 2009, chapters 7 and 9). Einstein also neglects to make it clear which investigations have revealed that his static solution is unstable, although this is very likely a reference to Eddington's paper on the subject (Eddington 1930; Nussbaumer 2014). These points are discussed further in section 3.

Einstein's second reason for ruling out his former static solution concerns the astronomical observations of Edwin Hubble:

> *"On the other hand, Hubbel's [sic] exceedingly important investigations have shown that the extragalactic nebulae have the following two properties:*
>
> 1) *Within the bounds of observational accuracy they are uniformly distributed in space*
> 2) *They possess a Doppler effect proportional to their distance"*

We note that Hubble's name is misspelt throughout the manuscript, as in the case of Einstein's *Sitzungsberichte* paper of 1931 (Einstein 1931b). This may indicate that Einstein was not fully familiar with Hubble's work, as has been previously argued (Nussbaumer 2014; O'Raifeartaigh and McCann 2014). We also note that Einstein uses the term 'Doppler effect' rather than radial velocity, suggesting a qualified acceptance of Hubble's observations as evidence for a cosmic expansion.

Remarking that the dynamic models of de Sitter and Tolman are consistent with Hubble's observations, Einstein points out that their models predict an age for the cosmos that is problematic:

> *"De Sitter and Tolman have already shown that there are solutions to equations (1) that can account for these observations. However the difficulty arose that the theory unvaryingly led to a beginning in time about $10^{10}$-$10^{11}$ years ago, which for various reasons seemed unacceptable."*

We note that there is again no reference to the evolving models of Friedman or Lemaître . The *"various reasons"* in the quote is almost certainly a reference to the fact that the estimated timespan of dynamic models was not larger than the ages of stars as estimated from astrophysics, or the age of the earth as estimated from radioactivity.[4] However, the sentence is a little ambiguous; it is possible that Einstein's "difficulty" also refers to the very notion of a *"beginning in time"* for the universe. Indeed, it is quite curious that the problem of origins is not specifically discussed in the manuscript (see section 3 below).

In the third part of the manuscript, Einstein explores an alternative solution to the field equations that could also be compatible with Hubble's observations – namely, an expanding universe in which the density of matter does not change over time:

> *"In what follows, I wish to draw attention to a solution to equation (1) that can account for Hubbel's [sic] facts, and in which the density is constant over time. While this solution is included in Tolman's general scheme, it does not appear to have been taken into consideration thus far."*

The reference to "Tolman's general scheme'' is significant as Einstein's ensuing analysis bears some technical similarities to a paper by Tolman in which the latter associated the cosmic expansion with a continuous transformation of matter into radiation (Tolman 1930).

Einstein starts the construction of his model by choosing the metric of flat space expanding exponentially:

> *"I let*
>
> $$ds^2 = -e^{\alpha t}(dx_1^2 + dx_2^2 + dx_3^2) + c^2 dt^2 \ldots \quad (2)$$
>
> *This manifold is spatially Euclidean. Measured by [this] yardstick, the distance between two points increases over time as $e^{\frac{\alpha}{2}t}$; one can thus account for Hubbel's Doppler effect by giving the masses (thought of as uniformly distributed) constant co-ordinates over time."*

Equation (2) represents the line element of the de Sitter universe in its simplest form.[5] A similar line element was employed by Tolman in the paper mentioned above (Tolman 1930) and Einstein's choice of metric may have been influenced by that work. However, it can be shown that the hypothesis of a constant rate of matter creation requires a metric that is spatially flat and exponentially expanding and Einstein may have realised this independently.[6]

---

[4] Einstein's view of the timescale problem is spelt out in detail in his later review of dynamic models (Einstein 1945).
[5] This was first shown by Robertson in 1928 (Robertson 1928).
[6] A constant rate of matter creation requires spatial flatness ($k = 0$) because the creation rate is related to spatial curvature ($k/R^2$) and the radius is not constant. A constant Hubble parameter $\dot{R}/R$ is also required, from which it follows that the expansion must be exponential.

Einstein notes that the metric is invariant:

> *"Finally, the metric of this manifold is constant over time. For it is transformed by applying the substitution*
>
> $$t' = t - \tau \quad (\tau = const)$$
>
> $$\frac{x'_1}{x_1} = \frac{x'_2}{x_2} = \frac{x'_3}{x_3} = e^{-\frac{\alpha}{2}\tau}$$
>
> *into*
>
> $$ds^2 = e^{\alpha t'}(dx_1'^2 + dx_2'^2 + dx_3'^2) + c^2 dt'^2 \ .$$

We note the apparent sign error in the last equation above, an error that may have led to a miscalculation in the derivation to follow.

Assuming a low velocity of masses relative to the co-ordinate system and negligible radiation pressure, Einstein constructs a matter-energy tensor in a manner analogous to his seminal paper of 1917 (Einstein 1917):

> *"We ignore the velocities of the masses relative to the co-ordinate system as well as the gravitational effect of the radiation pressure. The matter tensor is then to be expressed in the form*
>
> $$T^{ik} = \rho u^i u^k \quad (u^i = \frac{dx^i}{ds})$$
>
> *or*
>
> $$T_{ik} = \rho u^\sigma u^\tau g_{\sigma i} g_{\tau k} \ , \qquad (3)$$
>
> *where* $\quad u^1 = u^2 = u^3 = 0; u^4 = \frac{1}{c} \ ."$

From equations (1) - (3), he derives two simultaneous equations and eliminates the cosmological constant to solve for the matter density:

> *" Equations (1) yield:*
>
> $$\frac{-3[9?]}{4}\alpha^2 + \lambda c^2 = 0$$
> $$\frac{3}{4}\alpha^2 - \lambda c^2 = \kappa \rho c^2$$
>
> *or*
>
> $$\alpha^2 = \frac{\kappa c^2}{3}\rho \qquad \dots \qquad (4)"$$

Thus, Einstein has derived an expression for the matter density $\rho$ in terms of the expansion co-efficient $\alpha$. However, we note an ambiguity regarding the coefficient of $\alpha^2$ in the first of the simultaneous equations (see figure 2). While a value of +9/4 for this coefficient is implied by equation (4), it appears to have been later amended to **-3**/4, a correction that leads to the null solution *ρ = 0* instead of equation (4), exposing a fundamental flaw in the model (see section 4).

Before amending the analysis above, Einstein concluded from equation (4) that the density of matter remains constant and is related to the cosmic expansion:

> *"The density is therefore constant and determines the expansion apart from its sign."*

Thus, Einstein associates the cosmic expansion with a continuous creation of matter, while Tolman suggested that the expansion was driven by a transformation of matter into radiation (Tolman 1930).

In the final part of the manuscript, Einstein proposes a physical mechanism to allow the density of matter remain constant in a universe of expanding radius - namely, the continuous formation of matter from empty space:

> *"If one considers a physically bounded volume, particles of matter will be continually leaving it. For the density to remain constant, new particles of matter must be continually formed within that volume from space."*

This proposal anticipates the 'creation field' of Fred Hoyle in some ways (see section 5 below). However, unlike Hoyle, Einstein has not introduced a term representing this 'creation' process into the field equations (1). Instead he loosely associates the continuous formation of matter with the cosmological constant, suggesting that the latter ensures that the conservation of energy is not violated:

> *"The conservation law is preserved in that, by setting the λ-term, space itself is not empty of energy; its validity is well-known to be guaranteed by equations (1)."*

Thus, in this model of the cosmos, Einstein proposes that the cosmological constant assigns an energy to empty space that is associated with the creation of matter. However, the model is fundamentally flawed because the lack of a specific term representing matter creation leads to the null solution $\rho = 0$. We suggest that Einstein recognized this problem on revision of the manuscript and set the model aside rather than pursue more contrived steady-state solutions, as discussed in section 6.

## 3. Historical remarks: dating the manuscript

The manuscript under discussion has been assigned the year 1931 by the Albert Einstein Archive. However, this dating is no longer certain as the manuscript was mistaken for a draft of a different paper until now (see introduction).

It is instructive to attempt to date the manuscript from its contents. The statement *"Hubbel's [sic] exceedingly important investigations have shown that the extra-galactic nebulae… possess a Doppler effect proportional to their distance"* (section 2) gives confidence that the manuscript was written after Hubble's seminal publication of 1929 (Hubble 1929) as it is very unlikely that Einstein knew of Hubble's observations before this date (Nussbaumer 2014). Indeed, it is generally thought that Einstein's interest in cosmology was rekindled by Hubble's observations of the recession of the galaxies, and by his three-month stay in the United States from December 1930 to March 1931. Much of this trip was spent at Caltech, and included a meeting with Edwin Hubble and other astronomers at the Mount Wilson Observatory (Nussbaumer and Bieri 2009, chapter 14; Bartusiak 2009, p251-256; Eisinger 2011, p 109-115). Press reports of seminars given by Einstein at Caltech certainly suggest that he viewed Hubble's observations as likely evidence for an expanding universe. For example, *The New York Times* reported Einstein as commenting that *"New observations by Hubble and Humason concerning the redshift of light in distant nebulae make the presumptions near that the general structure of the universe is not static"* (AP 1931a) and *"The redshift of the distant nebulae have smashed my old construction like a hammer blow"* (AP 1931b).

Einstein had many interactions with the theoretician Richard Tolman at Caltech and greatly admired Tolman's work on relativity (Nussbaumer 2014; Nussbaumer and Bieri 2009, p145-147; Eisinger 2011, p114). Thus, Einstein may have been influenced by Tolman's cosmology; as noted in section 2, the manuscript under discussion bears some similarities to Tolman's 'annihilation' model of 1930 (Tolman 1930). Finally, Einstein's manuscript is written on American paper,[7] making it unlikely that it was written before his arrival in the United States in January 1931.

As regards an upper bound for the date of the manuscript, we note that it is very different in both style and content to the cosmic models published by Einstein in April 1931 (Einstein 1931b)[8] and 1932 (Einstein and de Sitter 1932). In the latter papers (known as the Friedman-Einstein and the Einstein-de Sitter models respectively), Einstein assumed that the mean density of matter varied with varying cosmic radius, and he removed the cosmological constant term from the field equations, pointing out that it was only introduced to keep the universe static. These 'evolving' models of the cosmos were not constructed from first principles, but employed the earlier analysis of Alexander Friedman; further, Einstein used Friedman's differential equations in conjunction with Hubble's observations to determine values for the radius and matter density of the universe (Einstein 1931b, Einstein and de Sitter 1932).

---

[7] We thank Barbara Wolff of the Albert Einstein Archives for confirming this point.
[8] It is known that this model was written in early April 1931 (see Eisinger 2011) chap 7.

By contrast, the model in the manuscript under discussion is constructed from first principles; the cosmological constant is not removed from the field equations and there is no reference to Friedman's analysis or to Einstein's evolving models of 1931 and 1932. We also note that there is no reference in the manuscript to the problem of cosmic origins articulated by Lemaître in 1931 (Lemaître 1931). Thus, it seems likely that the manuscript precedes the Friedman-Einstein and Einstein-de Sitter models, i.e., was written sometime in early 1931, and represents Einstein's first attempt at a cosmic model in the wake of emerging evidence for a cosmic expansion on the largest scales.[9]

## 4. Why was Einstein's steady-state model not published?

We note first that there is no mention of steady-state solutions in Einstein's later discussions of cosmic models (Einstein 1931b, Einstein and de Sitter 1932, Einstein 1945), nor have we been able to find a reference to the manuscript under discussion in Einstein's letters, diaries or other personal papers. This apparent silence indicates that he decided against the model, rather than simply mislaid the manuscript during his travels or neglected to publish it.

We suggest that Einstein's steady-state model was abandoned because it contains a fundamental flaw. In particular, we have been unable to reproduce Einstein's derivation of equation (4) from equations (1) - (3). Attempting to derive the simultaneous equations from first principles, we find a value of -3/4 for the coefficient of $\alpha^2$ in the first of the simultaneous equations, leading to the null solution $\rho = 0$ instead of equation (4). Close scrutiny of figure 2 suggests that the co-efficient of $\alpha^2$ in the first equation has indeed been amended to **-3**/4; it seems that Einstein discovered his initial error at a later point, realised the model led to a trivial solution and set the work aside without correcting equation (4).[10]

---
[9] We should emphasise that, in the absence of any definitive documentary evidence, this dating cannot be certain.

With modern eyes, it is easy to see why Einstein's steady-state model leads to a null solution; the fundamental problem is that he has not included a term in the analysis representing the hypothesised creation of matter. This leads one to ask why Einstein did not attempt a more sophisticated steady-state solution; we suggest that he may have decided that such an approach was overly contrived in comparison with evolving models, as discussed in section 6.

## 5. On steady-state models of the cosmos

We note first that diverse models of a 'steady-state' universe were advanced throughout the 20th century. In 1918, the American physicist William MacMillan proposed a continuous creation of matter from radiation in order to avoid a gradual 'running down' of the cosmos due to the conversion of matter into energy in stellar processes (MacMillan 1918, 1925). MacMillan's proposal was enthusiastically received by Robert Millikan, who suggested that the process might be the origin of cosmic rays (Millikan 1928). The idea of a continuous creation of matter from radiation was also considered by Richard Tolman as a means of introducing matter into the empty de Sitter universe, although he saw the idea as rather improbable (Tolman 1929).

Other physicists considered the possibility of the creation of matter from empty space. In 1928, James Jeans speculated that matter was continuously created in the centre of the spiral nebulae: *"The type of conjecture which presents itself, somewhat insistently, is that the centres of the nebulae are of the nature of "singular points", at which matter is poured into our universe from some other spatial dimension….so that they appear as points at which matter is poured into our universe from some other, and entirely extraneous spatial dimension, so that, to a denizen of our universe, they appear as points at which matter is continually created"* (Jeans 1928, p360). Similar ideas of continuous creation were explored by the Swedish scientist Svante Arrhenius and the

---

[10] Attempting to reconstruct Einstein's analysis, we take $T_{44} = \rho c^2$ (all other components zero) and from the time component of equation (1) we obtain $\left(R_{44} - \frac{1}{2}g_{44}R\right) - \lambda g_{44} = \kappa \rho c^2$. On analysis, this gives $-3\alpha^2/4 + 3\alpha^2/2 - \lambda c^2 = \kappa \rho c^2$, the second of Einstein's simultaneous equations. From the spatial component of equation (1), we obtain $\left(R_{ii} - \frac{1}{2}g_{ii}R\right) - \lambda g_{ii} = 0$, which on analysis gives $3\alpha^2/4 - 3\alpha^2/2 + \lambda c^2 = 0$ for the first of the simultaneous equations. It is plausible that Einstein made a sign error here, initially getting $3\alpha^2/4 + 3\alpha^2/2 + \lambda c^2 = 0$ for this equation.

German chemist Walther Nernst (Arrhenius 1908, 1909; Nernst 1928). However, these theories did not concern the creation of matter in an expanding universe.[11]

The concept of an expanding universe that remains in a steady state due to a continuous creation of matter is most strongly associated with the Cambridge physicists Fred Hoyle, Hermann Bondi and Thomas Gold. In the late 1940s, these scientists became sceptical of the evolutionary, Friedman-like models of the cosmos that had been proposed in the wake of Hubble's observations. They disliked Lemaître's idea of an explosive beginning for the universe (Lemaître 1931) and noted that the evolving models predicted an age for the cosmos that was problematic. They were also concerned that evolutionary models necessitated speculations about physical processes in the distant past that could not be tested directly (Hoyle 1948; Bondi 1952, chapter 12; Narlikar 1988, chapter 7). In order to circumvent these difficulties, the Cambridge trio explored the idea of an expanding universe that does not evolve over time, i.e., a cosmos in which the mean density of matter is maintained constant by a continuous creation of matter from the vacuum (Hoyle 1948; Bondi and Gold 1948).

In the case of Bondi and Gold, the proposal of a steady-state cosmos followed from their belief in the 'perfect cosmological principle', a philosophical principle that proposed that the universe should appear essentially the same to observers in all places *at all times*. This principle led them to postulate a continuous creation of matter in order to sustain an unchanging universe. While the idea bears some similarity to Einstein's attempt at a steady-state theory, it is difficult to compare the models directly because the Bondi-Gold theory was not formulated in the context of general relativity.[12]

On the other hand, Fred Hoyle constructed a steady-state model of the cosmos by means of a daring modification of the Einstein field equations (Hoyle 1948, Mitton 2005, chapter 5). Replacing Einstein's cosmological constant with a scalar field $C_{ik}$, representing the continuous creation of matter from the vacuum, Hoyle obtained the equation

$$\left(R_{ik} - \frac{1}{2}g_{ik}R\right) + C_{ik} = -\kappa T_{ik} \quad (5)$$

In this model, the expansion of space was driven by the creation of matter and the perfect cosmological principle emerged as a consequence rather than a starting assumption (Hoyle 1948).[13] In its initial form, Hoyle's model violated the principle of conservation of energy, but a more sophisticated version was advanced in later years (Hoyle and Narlikar 1962). In the latter model, matter creation was achieved via the introduction of negative stresses in the energy-momentum tensor on the right-hand side of the field equations.[14] An elegant mathematical formulation of the theory was then achieved using the principle of least action, as suggested by the Cambridge physicist Maurice Pryce (Hoyle and Narlikar 1962).

---

[11] See (Kragh 1996) p143-162 for a review of steady-state cosmologies in the early 20th century.
[12] The Bondi-Gold model was based on kinematic relativity because they were not convinced that general relativity could be applied to the cosmos on the largest scales (Bondi 1952, p145-146).

The steady-state model attempted by Einstein in the manuscript under discussion anticipates that of Hoyle in some respects, but the crucial difference is that Einstein did not introduce a term representing the creation of matter to the field equations (1), either on the left-hand side in the manner of Hoyle's initial model (Hoyle 1948) or on the right-hand side in the manner of the later Hoyle-Narlikar model (Hoyle and Narlikar 1962). This key omission led directly to the failure of Einstein's model; as discussed in section 6, it is interesting that he declined to explore more sophisticated steady-state models.

As is well known, a significant debate developed between steady-state and evolving theories of the cosmos during the 1950s and 1960s (Kragh 1996, chapter 5; Mitton 2005, chapter 7). Eventually, steady-state models were ruled out by astronomical observations that showed unequivocally that we inhabit a universe that is evolving over time.[15] We note that there is no evidence to suggest that any of the steady-state theorists above were aware of the manuscript under discussion; no doubt they would have been greatly interested to know that Einstein once considered a steady-state model of the cosmos.

## 6. On Einstein's philosophy of cosmology

It should come as no great surprise that, when confronted with empirical evidence for an expanding universe, Einstein once considered a stationary or steady-state model of the expanding cosmos. There is a great deal of evidence that Einstein's philosophical preference was for an unchanging universe, from his introduction of the cosmological constant to the field equations in 1917 to keep the universe static[16] to his well-known hostility to the dynamic models of Friedman and Lemaître when they were first suggested (Nussbaumer and Bieri 2009, chapters 7 and 9). Indeed, a model of an expanding cosmos in which the density of matter remains unchanged seems a natural successor to Einstein's static model of 1917, from a philosophical point of view.

---

[13] We note in passing that Hoyle also demonstrated that the model required a line element of the de Sitter form.
[14] This approach was first suggested by W. H. McCrea (McCrea 1951).
[15] The key results were the observation of the distribution of galaxies in the distant past and the discovery of the cosmic microwave background (Kragh 1996, chaper 7).
[16] While Einstein cited the low velocities of the stars in his static model (Einstein 1917), his choice was as much philosophical as empirical because there was no guarantee that a cosmic expansion would be detectable by astronomy.

However, such a steady-state theory requires the assumption of a continuous creation of matter and, as Einstein discovered, a successful model of the latter process was not possible without amending the field equations. On the other hand, an expanding universe of varying matter density could be described without any such amendment – and indeed without the cosmological constant, as Einstein proposed in the Friedman-Einstein and Einstein-de Sitter models (Einstein 1931b; Einstein and de Sitter 1932). It is therefore very probable that Einstein decided against steady-state models of the cosmos because they were more contrived than evolutionary models. This suggestion fits very well with our view of Einstein's minimalist approach to cosmology in these years.[17]

It is also possible that Einstein decided against steady-state models on empirical grounds, i.e., on the grounds that there was no observational evidence to support the postulate of a continuous formation of matter from empty space. It is interesting that, when asked to comment on Hoyle's steady-state model in later years, Einstein is reported to have dismissed the theory as "*romantic speculation*" (Michelmore 1962, p253). This criticism is confirmed in a letter written by Einstein to the physicist Jean Jacques Fehr in 1952. Einstein seems highly sceptical of Hoyle's model, and in particular of the postulate of a continuous creation of matter from the vacuum: "*Die kosmologischen Spekulationen von Herrn Hoyle, welche eine Entstehung von Atomen aus dem Raum voraussetzen, sind nach meiner Ansicht viel zu wenig begründet, um ernst genommen zu werden*" or "*The cosmological speculations of Mr Hoyle, which presume the creation of atoms from space, are in my view much too poorly grounded to be taken seriously*" (Einstein 1952).[18]

As pointed out in section 5, steady-state models of the cosmos were eventually ruled out by astronomical observation. Nevertheless, the model of this manuscript is of some interest in the study of Einstein's cosmology. In the first instance, it is significant that Einstein retained the cosmological constant in at least one cosmic model he proposed *after*

---

[17] Two of us have previously argued that Einstein's removal of the cosmological constant in 1931, followed by his removal of spatial curvature in 1932, suggests an Occam's razor approach to cosmology (O'Raifeartaigh and McCann, 2014).

[18] We thank Barbara Wolff of the Albert Einstein Archives for bringing this document to our attention.

Hubble's observations; it seems that the widely-held view that Einstein was happy to discard the cosmological constant at the first sign of evidence for a non-static universe (North 1965, p132; Kragh 1999, p34; Nussbaumer and Bieri 2009, p147) is not entirely accurate. Instead, it appears that Einstein's attraction to an unchanging universe at first outweighed his dislike of the cosmological constant, just as it did in 1917. Second, while it could be argued that Einstein's attempt at a steady-state theory was trivial because it didn't work, it is interesting that this was not evident to Einstein on first approach; it seems the de Sitter metric remained a source of some confusion at this point. It is even more interesting that, when his model failed, Einstein did not explore a more sophisticated steady-state theory that would have involved an amendment to the field equations but turned his attention to evolutionary models instead. Thirdly, Einstein's manuscript reminds us that today's model of an evolving cosmos did not occur as a sudden 'paradigm shift' in the wake of Hubble's observations. Instead, physicists explored diverse cosmic models for many years: from the possibility of an expansion caused by a continuous annihilation of matter (Tolman 1930) to one caused by condensation processes (Eddington 1930), from the conjecture that the redshifts of the nebulae represented a loss of energy by photons (Zwicky 1929) to the hypothesis of a steady-state universe (this manuscript).

We note finally that Einstein's attempted steady-state model has many points of interest for today's cosmologist. His association of the cosmological constant with an energy of space ("*by setting the λ-term, space itself is not empty of energy*") finds new relevance in the context of the recent discovery of an accelerated expansion and the postulate of dark energy. Indeed, many of today's models of dark energy bear echoes of steady-state theory; in particular, the recent hypothesis of 'phantom fields' has many points in common with the Hoyle-Narlikar model discussed in section 5 (Singh et al. 2003). In addition, the metric of an exponentially expanding, flat space employed by Einstein (and Hoyle) is precisely that used in modern theories of cosmic inflation. These points remind us of the relevance of past models of the universe for cosmology today.

## 7. Concluding remarks

The manuscript "*Zum kosmologischen Problem*" discussed in this paper was never formally published in the literature but is of historical interest because it features a model of the cosmos distinct from Einstein's static model of 1917 (Einstein 1917) or his evolving models of 1931 and 1932 (Einstein 1931b, Einstein and de Sitter 1932). In the manuscript,

Einstein considers a steady-state universe of expanding radius, non-zero cosmological constant and constant matter density, where the latter is maintained by the continuous formation of matter from space; this model anticipates the well-known steady-state cosmology of Hoyle, Bondi and Gold.

The manuscript was almost certainly written after 1929 and probably before Einstein's published cosmic models of 1931 and 1932. It was very likely abandoned because of a fundamental flaw in the specific model proposed; we suggest that Einstein declined to try again with a more sophisticated version because he came to the view that steady-state models were overly contrived. The manuscript is nevertheless of interest in the study of 20$^{th}$ century cosmology because it demonstrates that Einstein conducted an internal debate between steady-state and evolving models of the universe decades before a similar debate took place in the wider cosmological community.

**Acknowledgements**

The authors would like to thank the Albert Einstein Foundation of the Hebrew University of Jerusalem for permission to publish our translation of the manuscript and the excerpts shown in figures 1 and 2. Cormac O'Raifeartaigh thanks Micheal O'Keeffe of Waterford Institute of Technology and Professor John Barrow of the Department of Applied Mathematics & Theoretical Physics at Cambridge University for helpful discussions. Simon Mitton wishes to thank St Edmund's College, Cambridge for the provision of research facilities.

**Note added in proof**

An article giving further details on the historical background of Einstein's manuscript has recently been published on the Physics ArXiv by Professor Harry Nussbaumer (http://arxiv.org/abs/1402.4099).

**Figure 1**

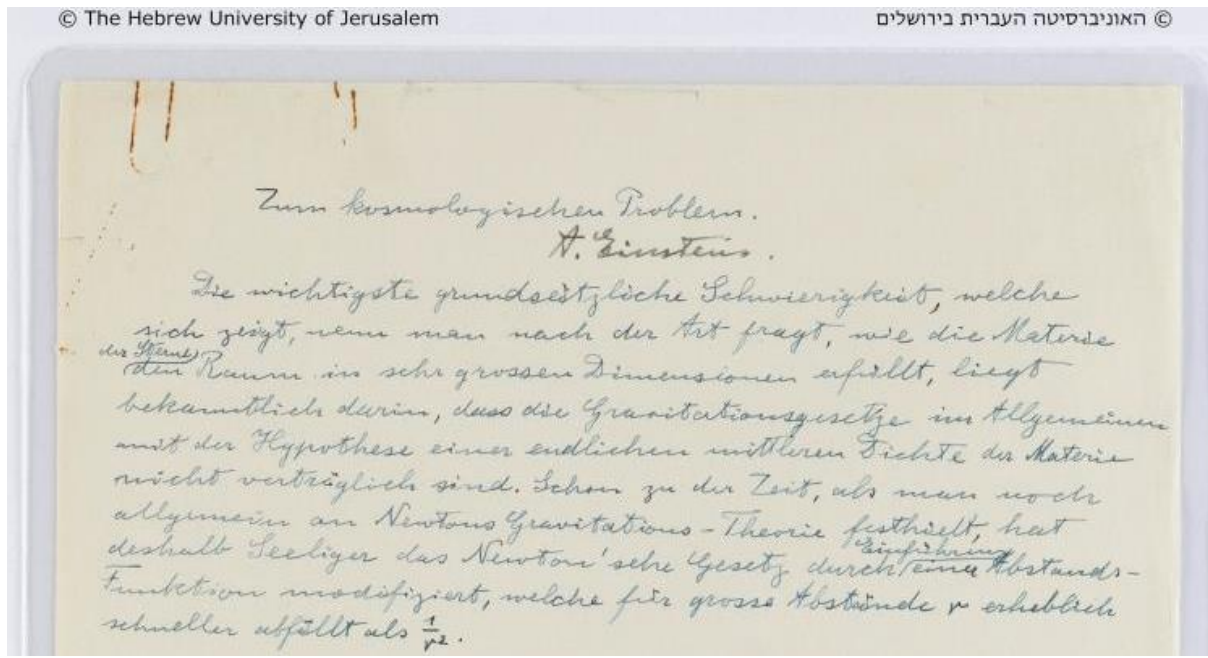

Title and opening paragraph of manuscript [2-112], reproduced from the Albert Einstein Archive by kind permission of the Hebrew University of Jerusalem.

**Figure 2**

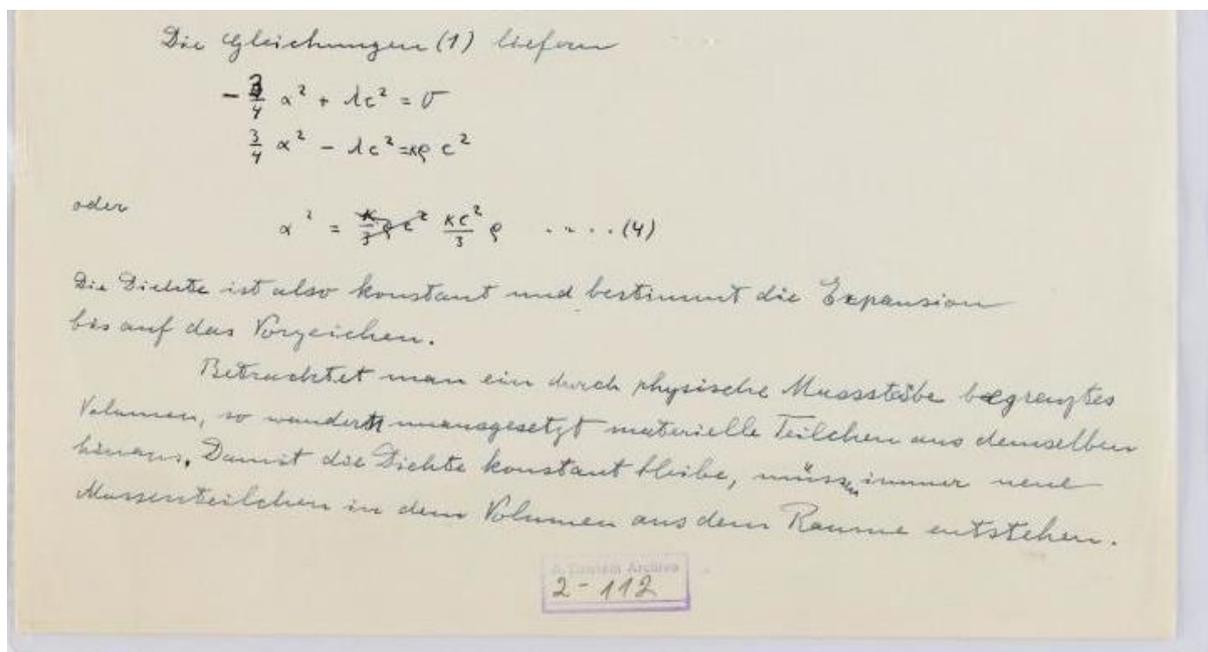

Excerpt from the third page of manuscript [2-112], reproduced from the Albert Einstein Archive by kind permission of the Hebrew University of Jerusalem. In the first of the simultaneous equations, the coefficient of the $\alpha^2$ term has been amended to **-3**/4 in different ink. The sentence following equation (4) states *"Die Dichte ist also konstant und bestimmt die Expansion bis auf das Vorzeichen"* or *"The density is therefore constant and determines the expansion apart from its sign"*

**Appendix**     **Translation of Einstein's manuscript**[ϕ]

On the cosmological problem

### A. Einstein

It is well known that the most important fundamental difficulty that emerges when one asks how the stellar matter fills up space in very large dimensions is that the laws of gravity are not in general consistent with the hypothesis of a finite mean density of matter. Thus, at a time when Newton's theory of gravity was still generally accepted, Seeliger had already modified the Newtonian law by the introduction of a distance function that, for large distances r, diminishes considerably faster than $1/r^2$.

This difficulty also arises in the general theory of relativity. However, I have shown in the past that this can be overcome by the introduction of the so-called "λ–term" to the field equations. The field equations can then be written in the form

$$\left(R_{ik} - \frac{1}{2}g_{ik}R\right) - \lambda g_{ik} = \kappa T_{ik} \quad \ldots (1)$$

$$R_{ik} = \Gamma^{\sigma}_{ik,\sigma} - \Gamma^{\sigma}_{i\sigma,k} - \Gamma^{\sigma}_{i\tau}\Gamma^{\tau}_{k\sigma} + \Gamma^{\sigma}_{ik}\Gamma^{\tau}_{\sigma\tau} \quad \ldots 1(a)$$

At that time, I showed that these equations can be satisfied by a spherical space of constant radius over time, in which matter has a density ρ that is constant over space and time.

It has since transpired that this solution is almost certainly ruled out for the theoretical comprehension of space as it really is.

On the one hand, it follows from investigations based on the same equations by [ ] and by Tolman that there also exist spherical solutions with a world radius P that is variable over time, and that my solution is not stable with respect to variations of P over time. On the other hand, Hubbel's exceedingly important investigations have shown that the extragalactic nebulae have the following two properties:

1) Within the bounds of observational accuracy they are uniformly distributed in space
2) They possess a Doppler effect proportional to their distance.

De Sitter and Tolman have already shown that there are solutions to equations (1) that can account for these observations. However the difficulty arose that the theory unvaryingly led to a beginning in time about $10^{10}$ -$10^{11}$ years ago, which for various reasons seemed unacceptable.

---

[ϕ] Translated from the original manuscript by Cormac O'Raifeartaigh and Brendan McCann by kind permission of the Hebrew University of Jerusalem. The manuscript can be viewed as Document [2-112] on the Albert Einstein Archive Online at http://alberteinstein.info/vufind1/Record/EAR000034354.

In what follows, I wish to draw attention to a solution to equation (1) that can account for Hubbel's facts, and in which the density is constant over time. While this solution is included in Tolman's general scheme, it does not appear to have been taken into consideration thus far.

I let

$$ds^2 = - e^{\alpha t}(dx_1^2 + dx_2^2 + dx_3^2) + c^2 dt^2 \ldots \qquad (2)$$

This manifold is spatially Euclidean. Measured by [this] yardstick, the distance between two points increases over time as $e^{\frac{\alpha}{2}t}$; one can thus account for Hubbel's Doppler effect by giving the masses (thought of as uniformly distributed) constant co-ordinates over time. Finally, the metric of this manifold is constant over time. For it is transformed by applying the substitution

$$t' = t - \tau \quad (\tau = const)$$

$$\frac{x_1'}{x_1} = \frac{x_2'}{x_2} = \frac{x_3'}{x_3} = e^{-\frac{\alpha}{2}\tau}$$

into

$$ds^2 = e^{\alpha t'}(dx_1'^2 + dx_2'^2 + dx_3'^2) + c^2 dt'^2 \ .$$

We ignore the velocities of the masses relative to the co-ordinate system as well as the gravitational effect of the radiation pressure. The matter tensor is then to be expressed in the form

$$T^{ik} = \rho u^i u^k \quad (u^i = \frac{dx^i}{ds})$$

or

$$T_{ik} = \rho u^\sigma u^\tau g_{\sigma i} g_{\tau k}, \qquad \ldots (3)$$

where
$$u^1 = u^2 = u^3 = 0; \ u^4 = \frac{1}{c} \ .$$

Equations (1) yield

$$\frac{-3[9?]}{4}\alpha^2 + \lambda c^2 = 0$$

$$\frac{3}{4}\alpha^2 - \lambda c^2 = \kappa \rho c^2$$

or

$$\alpha^2 = \frac{\kappa c^2}{3}\rho \qquad \ldots \ (4)$$

The density is therefore constant and determines the expansion apart from its sign.

If one considers a physically bounded volume, particles of matter will be continually leaving it. For the density to remain constant, new particles of matter must be continually formed within that volume from space. The conservation law is preserved in that, by setting the λ-term, space itself is not empty of energy; its validity is well known to be guaranteed by equations (1).

## **Typographical notes from the translators**

(i) We have preserved the layout of the original manuscript in terms of sentence structure, paragraph structure and numbering of equations.
(ii) The name Hubble is misspelt as Hubbel each time it occurs in the manuscript. We have retained Einstein's misspelling as it may be of historical significance (see section 2).
 (iii) Some words and terms were crossed out in the manuscript; these deletions are not reproduced in our translation.
(iv) Entities in rectangular brackets indicate omissions in the original manuscript. In the fifth paragraph, the empty bracket [ ] indicates a missing reference. In the sentence below equation (2), [this] denotes a missing word in the original German. In the simultaneous equations above equation (4), th symbol [-9?] in the first equation denotes an apparent amendment by Einstein of the coefficient of the $α^2$ term from 9/4 to **-3**/4 (see section 4).